\documentclass[aip]{revtex4-2}

\usepackage{romannum}
\usepackage{graphicx}
\usepackage{amsmath}
\usepackage{hyperref}
\usepackage{epstopdf}

\hypersetup{
    colorlinks = true,
    allcolors = {blue},
}
\newcommand{\lc}{\textsubscript}

\newcommand{\RomanNumeralCaps}[1]
    {\MakeUppercase{\romannumeral #1}}
\begin{document}

\title{Coherent phonon-magnon interactions detected by micro-focused Brillouin light scattering spectroscopy}%

\author{Yannik Kunz} 
\email{ykunz@rptu.de}
\affiliation{Fachbereich Physik and Landesforschungszentrum OPTIMAS, Rheinland-Pfälzische Technische Universität Kaiserslautern-Landau, 67663 Kaiserslautern, Germany}
\author{Matthias Küß}
\affiliation{Institute of Physics, University of Augsburg, 86135 Augsburg, Germany}
\author{Michael Schneider}
\author{Moritz Geilen}
\author{Philipp Pirro}
\affiliation{Fachbereich Physik and Landesforschungszentrum OPTIMAS, Rheinland-Pfälzische Technische Universität Kaiserslautern-Landau, 67663 Kaiserslautern, Germany}
\author{Manfred Albrecht}
\affiliation{Institute of Physics, University of Augsburg, 86135 Augsburg, Germany}
\author{Mathias Weiler}
\affiliation{Fachbereich Physik and Landesforschungszentrum OPTIMAS, Rheinland-Pfälzische Technische Universität Kaiserslautern-Landau, 67663 Kaiserslautern, Germany}

\date{\today}%

\begin{abstract}
We investigated the interaction of surface acoustic waves and spin waves with spatial resolution by micro-focused Brillouin light scattering spectroscopy in a Co\lc{40}Fe\lc{40}B\lc{20} ferromagnetic layer on a LiNbO\lc{3}-piezoelectric substrate. We experimentally demonstrate that the magnetoelastic excitation of magnons by phonons is coherent by studying the interfering BLS-signals of the phonons and magnons during their conversion process.
% studying the interference of light scattered off generated magnons and annihilated phonons. 
We find a pronounced spatial dependence of the phonon annihilation and magnon excitation which we map as a function of the magnetic field. The coupling efficiency of the surface acoustic waves (SAWs) and the spin waves (SWs) is characterized by a magnetic field dependent decay of the SAWs amplitude.  
\end{abstract}

\maketitle

%%%%%%%%%%%%%%%%%%%%%%%%%%%%%%%%%%%%%%%%%%%%%%%%%%%%%%%%%%%%%%%%%%%%%%%%%%%%%%%%%%%%%%%%%%%%%%%%%%%%%%%%%%%%%%% Introduction %%%%%%%%%%%%%%%%%%%%%%%%%%%%%%%%%%%

Surface Acoustic Waves (SAW) with frequencies in the gigahertz regime have wavelengths on the micrometer scale. They thus enable the  miniaturization of microwave components and are ubiquitous in everyday devices~\cite{Campbell1998, Laenge2008, Franke2009}. SAW devices are further used for instance for probing material properties,~\cite{Lomonosov2007}, rf signal processing~\cite{Viktorov1967, Hashimoto2000} or sensors~\cite{Paschke2017}. Interdigital transducers (IDTs) thereby enable coherent and energy-efficient excitation and detection of SAWs on piezoelectric substrates with sufficiently small insertion losses for quantum applications~\cite{Ekstrom:Surface:2017}.
%~\cite{Geilen2020} provide references for SAWs here!
If SAWs propagate in magnetically ordered materials, the coupling of acoustic and magnetic excitations opens up a wide branch of possibilities~\cite{Yang2021, Kuss:Chiral:2022}. The magnetoacoustic control enables for instance magnetic switching~\cite{Li2014, Thevenard:Precessional:2016}, the creation and control of skyrmions~\cite{Tomoyuki2020, Chen:Ordered:2023}, the generation of Terahertz radiation~\cite{Zhang2020}, magnetic field controlled phase-shifting of acoustic waves~\cite{Rovillain2022}, acoustically driven linear and non-linear spin-wave resonance~\cite{Dreher2012, Weiler2011,Geilen2022_2, Shah:Symmetry:2023}, and acoustic spin-charge conversion~\cite{Weiler:Spin:2012, Kawada:Acoustic:2021}. The coupling of SAWs and spin waves (SW) breaks time-reversal symmetry, and the concomitant non-reciprocal SAW transmission~\cite{Xu2020, Kuess2020, Kuss:Giant:2023} may find applications for non-reciprocal miniaturized microwave devices~\cite{Verba2018, Liang2009, Verba2021}.\\
Commonly, the interaction between SAWs and SWs devices is studied using electrical measurement techniques by determining the magnetic field-dependent SAW transmission from IDT to IDT as detailed, e.g., in~\cite{Kuess2020, Weiler2011}. Measuring the SAW transmission allows for studying the SW dispersion and the symmetry of the magnetoacoustic interaction~\cite{Kuess2021_2}. However, this electrical measurement technique does not offer spatial resolution. While the widely used model for SAW-SW interaction~\cite{Dreher2012} implicitly assumes coherent SAW-SW interaction as the mechanisms causing the detected SAW absorption, experimental proof for the coherency is missing. Previous studies used imaging techniques to resolve SAW propagation in magnetic media~\cite{Zhao:Direct:2021,Casals:Generation:2020, Kraimia:Time:2020} and established separate detection of SAW and SW signals~\cite{Geilen:Fully:2022}. However, these works could not demonstrate the spatial dependency of the SAW-SW conversion, and the coherency of the SAW and SW remains an additional important open question as identified in Ref.~\cite{Casals:Generation:2020}. 

Here, we use microfocused Brillouin light scattering (µBLS) to study the magnetoacoustic interaction of SAWs with SWs on a LiNbO\lc{3}/Co\lc{40}Fe\lc{40}B\lc{20}(10\;nm)-structure with frequency- and spatial resolution. By taking advantage of the tunable sensitivity of µBLS to both phonons and magnons~\cite{Kargar2021, Geilen:Fully:2022}, we are able to separately investigate the absorption of phonons and the excitation of magnons in the system. We observe clear experimental evidence for the coherence of annihilated phonons and generated magnons by interference of the two corresponding signals, which leads to a distortion of the typical Lorentzian lineshape. This results in a Fano-resonance-like lineshape~\cite{Fano1961, Yong:2006:ClassFano} as predicted for magnetoacoustic waves by Latcham \textit{et al.}~\cite{Latcham2019}. We further reveal the spatial dependency of the phonon-magnon conversion process within the 10\;nm thick Co\lc{40}Fe\lc{40}B\lc{20} (CoFeB) film. A schematic depiction of the used µBLS-setup is shown in Fig.\;\ref{Fig:Setup+IDT_Spectra} a). A more detailed description of the setup is given in the supplementary material.
    
%%%%%%%%%%%%%%%%%%%%%%%%%%%%%%%%%%%%%%%%%%%%%%%%%%%%%%%%%%%%%%%%%%%%%%%%%%%%%%%%%%%%%%%%%%%%%%%%%%%%%%%%%%%%%%% Results %%%%%%%%%%%%%%%%%%%%%%%%%%%%%%%%%%%%%%%%
\begin{figure}
\includegraphics[scale = 1]{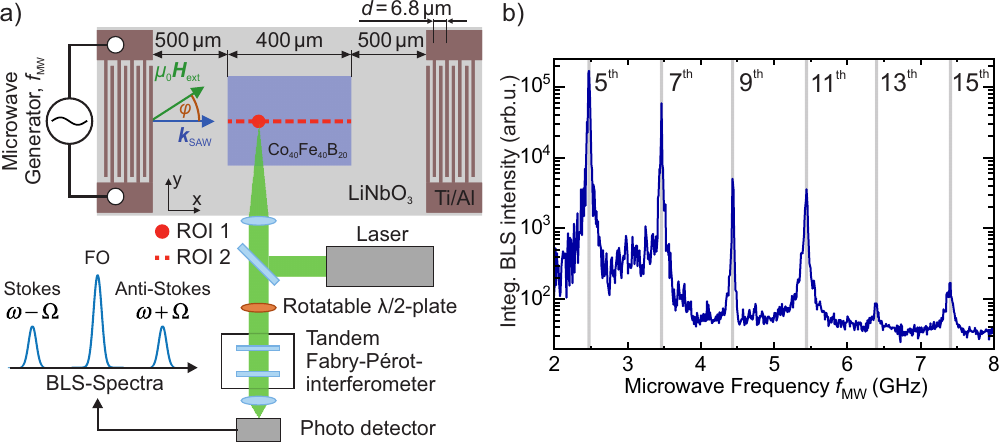}
\caption{Panel a) Schematic depiction of the used sample and the measurement configuration. On a LiNbO\lc{3} piezoelectric substrate, a 10\;nm thick and 400\;µm wide Co\lc{40}Fe\lc{40}B\lc{20} layer is deposited between two sets of IDTs with a finger periodicity of 6.8\;µm and 30 finger pairs (not all shown in the figure). The external magnetic field $\mu_0H_{\mathrm{ext}}$ is oriented along $\varphi\approx32.6\;^{\circ}$ relative to the propagation direction of the SAW $k_{\mathrm{SAW}}$. We used microfocused BLS for phonon and magnon spectroscopy, while a microscope camera allows for measuring with space resolution (not shown). The position of the laser spot during the measurements is indicated by the red dot (fixed position) and the red dashed line (linescan). The ROI 2 starts at the beginning of the ferromagnetic layer. Panel b) The excitation spectra of the employed set of IDTs is determined by integration of the detected BLS intensity. The excitation peaks arise if the condition of constructive interference for emitted SAWs between the IDT fingers is fulfilled.}
\label{Fig:Setup+IDT_Spectra}
\end{figure}
In the first part of our investigation, we characterized the phonon-spectra excited by the IDT by varying the applied rf-frequency $f_\text{MW}$ in the range of 2 to 8\;GHz. The obtained BLS-spectra were integrated in BLS-frequency for each rf-frequency. The resulting phonon excitation spectrum of the IDT is shown in Fig.\;\ref{Fig:Setup+IDT_Spectra} b). Excitation peaks arise periodically at frequencies $f_n$, fulfilling the condition of constructive interference
\begin{equation}
f_n=\frac{c_{\mathrm{SAW}}}{d}n\approx500\;\text{MHz}\cdot n,\hspace{0.5cm}n\in\{1, 3, 5,...\},
\end{equation}
where $d$ denotes the periodicity of the IDT. In Fig.\;\ref{Fig:Setup+IDT_Spectra} b) the lowest frequency peak corresponds to the 5\textsuperscript{th} harmonic order of the IDT at $2.48$\;GHz.\\
\begin{figure}
\includegraphics[scale = 1]{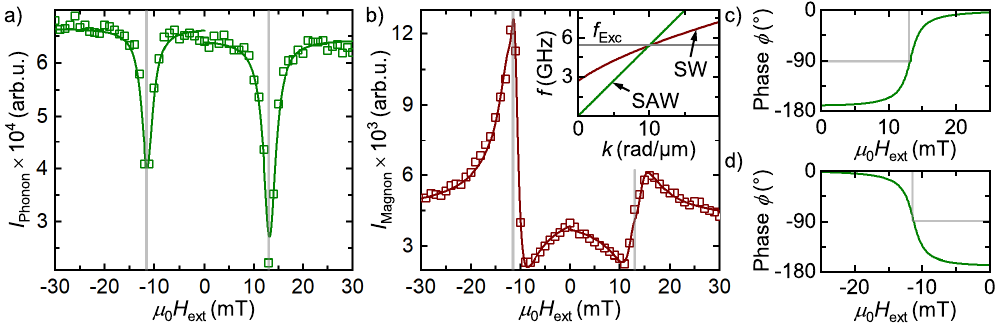}
\caption{Panel a) shows the integrated BLS-intensity at ROI 1 measured on phonon-polarization as a function of the external magnetic field $\mu_0H_{\mathrm{ext}}$. Two dips form at the positive and negative resonant magnetic field (gray lines) with different magnitudes, indicating nonreciprocal coupling. Squares denote the experimental data and solid curves the fit, in panel a) according to Eq.\;\eqref{eq:Iphonon} and in panel b) to Eq.\;\eqref{Eq:MagnonFitFunc}. In b) the resulting BLS-intensity close to pure magnon-polarization is shown. The phase shift between generated magnons and annihilated phonons affects the detection via µBLS and leads to dip-peak-like behavior. The inset in b) shows the triple crosspoint between the excitation frequency $f_{\mathrm{Exc}}=5.45$\;GHz and the dispersion relations of the SAW and the SWs at $\mu_0H=11$\;mT. In c) and d) the phase shift $\phi$ between the phonons and the magnons as a function of the external magnetic field is shown.}
\label{Fig:BLS_HSweep}
\end{figure}
To investigate the magnetic field-dependent coupling of phonons and magnons, the laser spot is positioned about 100\;µm into the ferromagnetic layer at "ROI 1", as indicated in Fig.\;\ref{Fig:Setup+IDT_Spectra} a). We make use of the rotatable $\lambda/2-$plate, which allows for tuning the relative sensitivity of our BLS setup to magnons or phonons \cite{Geilen:Fully:2022}. We excited the SAW at a rf-frequency of 5.45\;GHz (11\textsuperscript{th} order) and microwave output power of +18\;dBm. The sample was oriented so that the angle between the propagation direction of the SAW given by $\mathbf{k}_{\mathrm{SAW}}$ and the external magnetic field $\mu_0H_{\mathrm{ext}}$ was about $\varphi\approx33^{\circ}$. We integrated the BLS-spectra in the range of -5.25\;GHz to -5.925\;GHz for both the phonon and the magnon polarization of the $\lambda/2-$plate. The resulting intensities as a function of the external magnetic field are given in Fig.\;\ref{Fig:BLS_HSweep} for a) the phonon- and b) the magnon signal.\\
First, we discuss in panel a) the phonon signal. Here, dips in the BLS signal are observed at a positive magnetic field of $\mu_0H_{\mathrm{ext}}=13$\;mT and a negative field of $\mu_0H_{\mathrm{ext}}=-11.5$\;mT. The concomitant reduction in phonon number is attributed to resonant magnon-phonon coupling at the triple crosspoint (see the inset in Fig.\;\ref{Fig:BLS_HSweep} b)) between the linear SAW dispersion relation $f_\text{SAW}=c_\text{SAW}k/2\pi$ (green), the SW dispersion relation (red), and the excitation frequency (grey). The magnon dispersion relation is calculated using the Kalinikos-Slavin-equation~\cite{KalinikosSlavin1986}
\begin{align}
    f_\text{SW}(k,H)=&\frac{g\mu_\mathrm{B}\mu_0}{2\pi\hbar}\cdot\sqrt{H_\text{ext}+\frac{2A}{M_\mathrm{S}}k^2+H_\mathrm{ani}+M_\mathrm{S}\cdot\left(\frac{1-e^{-\|k\|t}}{\|k\|t}\right)}\notag\\
    &\cdot\sqrt{H_\text{ext}+\frac{2A}{M_\mathrm{S}}k^2+H_\mathrm{ani}+M_\mathrm{S}\cdot\left(1-\frac{1-e^{-\|k\|t}}{\|k\|t}\right)\sin^2(\varphi)}.
    \label{Eq:KS_Eqa}
\end{align}
We used broadband ferromagnetic resonance spectroscopy (see supplementary material) to determine the g-factor $g=2.11(8)$, Gilbert damping parameter $\alpha=0.006(7)$, saturation magnetization $\mu_0M_{\mathrm{S}}=1.287$\;T, and anisotropy field $\mu_0H_{\mathrm{ani}}=-1.46$\;mT of our CoFeB film. The small field shift between the positive and negative resonance magnetic field is attributed to an offset of the Hall probe rather than any SW nonreciprocity.
The different dip intensities for positive and negative magnetic fields are attributed to the helicity mismatch effect~\cite{Xu2020, Kuess2020}. When changing directions of the magnetic field, the helicity of the spin wave is inverted, while the helicity of the SAW remains the same, as it is determined by the SAWs propagation direction. The helicity mismatch effect gives rise to different coupling efficiencies on whether the helicities match (pos. field) or mismatch (neg. field), thus leading to different dip magnitudes~\cite{Kuss:Chiral:2022}.\\ 
We model the obtained phonon signal as follows. Following Ref.~\cite{Cardona1982}, we assume that the obtained phonon BLS intensity is proportional to the out-of-plane displacement $u_z^2$
\begin{equation}
    I_{\text{Ph}}(x,H)\propto \int\limits_0^{k_\text{max}}\int\limits_0^Tu_z^2(x,t,H)\text{d}t\text{d}k.
\end{equation}
The displacement after a certain propagation distance $x$ becomes magnetic field dependent due to the absorption of SAW phonons by SW generation. We derive the absorption of SAW power by following the approach of Küß \textit{et al.}~\cite{Kuess2020} and make use of the correlation between the SAW power and the displacement $P_{\text{SAW}}\propto u_z^2$. Thus, the expected BLS intensity can be written as
\begin{equation}
\label{eq:Iphonon}
    I_{\text{Ph}}(x,H)= I_0\cdot\exp{(-C_1\text{Im}(\chi_{11}(H))x)},
\end{equation}
where $I_0$ is the BLS intensity obtained from the SAW at the launching IDT, $C_1$ is a constant that quantifies the SAW-SW conversion efficiency and $\chi_{11}(H)$ is the diagonal component of the magnetic susceptibility tensor $\chi$. In this simple model, the SAW-SW helicity mismatch effect is phenomenologically taken into account by using different $C_1$ for $\mu_0H_\text{ext}<0$ and $\mu_0H_\text{ext}>0$. We derive $\chi$ by solving the Landau-Lifshitz-Gilbert-equation (see supplementary material) and fit Eq.\;\eqref{eq:Iphonon} to the data in Fig.\;\ref{Fig:BLS_HSweep} a) with fitting parameters $I_0$ and $C_1$. As can be seen in Fig.\;\ref{Fig:BLS_HSweep}\;a), good agreement between the BLS-intensity and the fitting model can be obtained.

Next, we consider the obtained BLS signal measured at optimized magnon detection efficiency by rotating the $\lambda/2-$plate correspondingly. Since the number of phonons is always significantly higher than the number of newly excited magnons, the fraction of the unfiltered phonon signal cannot be neglected and has to be taken into account. We observe a peak-dip-like behavior, as can be seen in Fig.\;\ref{Fig:BLS_HSweep} b), which we explain as follows: On resonance, phonons are annihilated, and magnons are generated. Because of the coherency of this process, the photons inelastically scattering off these magnons and phonons can interfere with each other. When sweeping through the resonance field, the phase relation between magnons and phonons changes, as detailed below, so that the interference is destructive/constructive depending on the magnetic field. This leads to a Fano-resonance-like lineshape~\cite{Fano1961}.
\begin{figure}
\includegraphics[width =\textwidth]{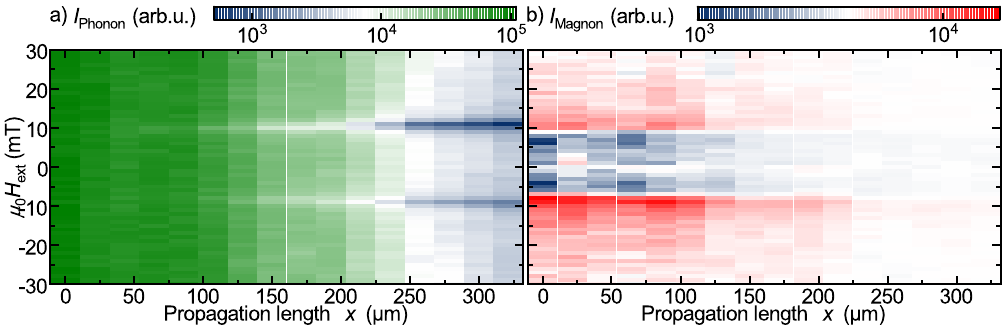}
\caption{Integrated BLS-intensity of the linescan measurement indicated by ROI 2 in Fig.\;\ref{Fig:Setup+IDT_Spectra} a), as a function of the external magnetic field and the propagation length of the SAW. In a) the measured intensity on phonon-polarization is shown, where it can be seen that with increasing propagation length two dips form at the resonant magnetic field. Panel b) shows the obtained intensity on magnon polarization (the scaled intensity on phonon-polarization at 30\;mT is subtracted). The highest increase in magnon population occurs at the start of the ferromagnetic layer at resonant magnetic field and decreases with vanishing phonon amplitude.} 
\label{Fig:Colormaps}
\end{figure}

\begin{figure}
\includegraphics[scale = 1]{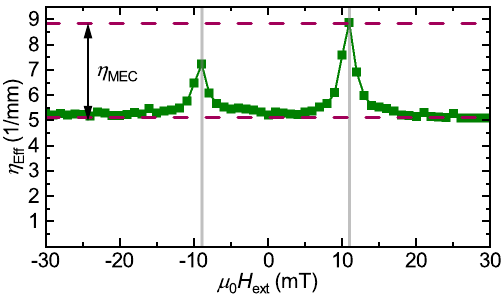}
\caption{Increase of the phonon decay rate due to the magnetoelastic coupling with SW at 5\;GHz as a function of the external magnetic field. The different magnitudes in peaks is denoted to the nonreciprocity inducing helicity mismatch effect.}
\label{Fig:AbsorptionRates}
\end{figure}

To describe the obtained signal, we start by writing the BLS intensity as
\begin{equation}
    I_{\text{Ma}}(x,H)\propto \int\limits_0^{k_\text{max}}\int\limits_0^T(c_\text{Ph}u_z(x,t,H)+c_\text{Ma}m_z(x,t,H))^2\text{d}t\text{d}k\label{Ima},
\end{equation}
with $c_\text{Ph}$ and $c_\text{ma}$ representing the detection efficiency of the phononic and the magnonic signal at the given position of the $\lambda/2-$plate, $m_z$ the dynamic out-of-plane magnetization component of the SW and $u_z$ the displacement due to the SAW. For the dynamic magnetization component $m_z$ and the displacement $u_z$ we make the generalized wave-like Ansatz
\begin{align}
    u_z(x,t,H)&=u_{z,0}(x,H)\cdot\exp(i(\omega t-kx)),\\
    m_z(x,t,H)&=m_{z,0}(x,H)\cdot\exp(i(\omega t-kx)-\phi(H))),
\end{align}
where $u_{z,0}(x,H)$ and $m_{z,0}(x,H)$ are the magnetic field and spatially dependent amplitudes of the displacement and the dynamic magnetization. We also include the phase shift $\phi(H)$ between the SAW driven dynamic magnetization and the SAW itself, similar to the classical driven harmonic oscillator, where a phase shift of $\phi=90^\circ$ is expected at resonance. The displacement $u_z$ can be derived as previously from the SAW power in Eq.\;\eqref{eq:Iphonon}.\\    
The magnetic component $m_z$ is derived by the locally absorbed SAW power as due to the high Gilbert damping in the system, only the locally excited magnons contribute to the BLS signal, as will be discussed in more detail later on. By assuming that the SW power is proportional to the dynamic out-of-plane magnetization component $P_\text{SW}\propto\overline{m_z^2}$ and the locally absorbed SAW power $P_\text{abs,loc}$ flows into the spin wave system, we obtain
\begin{equation}
    m_{z}(x,H)\propto\sqrt{\text{Im}(\chi_{11})}h_\text{dr}(x,H),
\end{equation}
 with the driving field $h_\text{dr}$ generated by the SAW. As the amplitude of the SAW decreases with increasing propagation length, so does the driving field. In turn, the driving field $h_\text{dr}$ can again be derived using the displacement $u_z$ by $h_\text{dr}\propto u_z$. Thus, we obtain for the expected BLS intensity by only taking the real part of Eq.\;\eqref{Ima} and neglecting higher order terms that are not linear in $c_\text{Ph}$
\begin{align}
    I_{\text{Ma}}(x,H)\propto&c_\text{Ma}^2C_2\text{Im}(\chi_{11})\exp(-C_1\text{Im}(\chi_{11})x)+\notag\\
    &2c_\text{Ph}c_\text{Ma}\sqrt{\text{Im}(\chi_{11})}\exp({-C_3\text{Im}(\chi_{11})x})\cos(\phi).\label{Eq:MagnonFitFunc}
\end{align}
Here, $C_2$ and $C_3$ are again constant prefactors included for simplification and to combine other constant prefactors. We use Eq.\;\eqref{Eq:MagnonFitFunc} to fit the data in Fig.\;\ref{Fig:BLS_HSweep}\;b). As can be seen, we achieve good agreement between the obtained experimental data and our model. In Fig.\;\ref{Fig:BLS_HSweep}\;c) and d) the resulting phase shift between the SAW and the SW is shown, becoming -90$^\circ$ at the resonant coupling field in agreement with the expectation for a driven harmonic oscillator. Thus, our experimental data provides evidence for a well-defined phase relation and thus coherency between the annihilated phonons and generated magnons.
Next, we map the magnetoelastic coupling as a function of the external magnetic field and the propagation distance of the SAW. For this, we use an excitation frequency of 2.48\;GHz at an excitation power of +18\;dBm and exploit the second-order harmonic generation~\cite{Lean1970, Lean1970_2, Kraimia:Time:2020} of the 10\textsuperscript{th} order IDT resonance at 5\;GHz to investigate the space-dependent coupling. The magnetic field was aligned as before ($\varphi\approx32.6^{\circ}$), however, now a linescan measurement was performed, as indicated by the red dashed line labeled "ROI 2" in Fig.\;\ref{Fig:Setup+IDT_Spectra} a). Again, we measured using both phonon and magnon polarization and integrated the resulting BLS-spectra in BLS frequency. The results are presented in Fig.\;\ref{Fig:Colormaps}, in panel a) for the obtained phonon signal and in b) the magnon signal, as a function of the applied magnetic field $\mu_0H_{\mathrm{ext}}$ and the propagation length $x$. Here, the scaled intensity on phonon-polarization at 30\;mT is subtracted in order to remove the unfiltered phononic signal. \\
First, we discuss the obtained phonon signal. Here, two dips of different magnitudes start to form with increasing propagation length $x$ over the ferromagnetic layer. The magnetic fields at which the dips occur again correspond to the triple crosspoint between the excitation frequency and the dispersion relations of the SAW and the SW, as discussed before. The magnetic field dependence of the phonon signal becomes more pronounced with increased SAW propagation because of the progressive SAW absorption during its propagation in the CoFeB film. This finding supports the previously observed dependence of the electrically detected magnetoelastic interaction on the length of the magnetic film~\cite{Kuss:Giant:2023}.\\
We now turn to the magnon signal shown in Fig.\;\ref{Fig:Colormaps} b), where the highest excitation of magnons is found at the beginning of the ferromagnetic layer. Due to the considerably large Gilbert damping $\alpha$ in the CoFeB film, magnons have a very low lifetime, leading to a short decay length $\xi_\text{SW}$ of only about $\xi_\text{SW}\approx$\;1.81\;µm at $f=5$\;GHz and $\mu_0H=10$\;mT in Damon-Eshbach geometry (see supplementary material), thus vanishing almost instantly. Consequently, the magnon population does not build up with increasing propagation length and only locally excited SWs by the SAW are detected. Since the phonon density is highest at the start of the ferromagnetic layer, the highest excitation of magnons is found here, while fewer magnons are exited with increasing propagation length. 
\begin{figure}
\includegraphics[scale = 1]{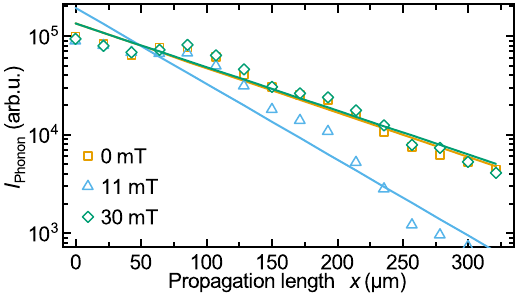}
\caption{Decrease of the SAW amplitude with propagation length, shown for different magnetic fields. At 11\;mT (resonant magnetic coupling field) the decrease in SAW amplitude is enhanced compared to off-resonant magnetic fields.}
\label{Fig:Spacedep_PhononAbs}
\end{figure}
The coupling of the phonon to the magnon system opens a loss channel for the propagating SAW phonons. From the previously obtained data, we now determine the magnetic field dependency of the SAW amplitude decay rate.  We obtain an exponential decrease in SAW amplitude with propagation distance which is characterized by the effective damping parameter $\eta_{\mathrm{eff}}(H)$, which we derive from the BLS-intensity by fitting
\begin{align}
I_{\mathrm{Phonon}}(x,H)=I_{0,\mathrm{Phonon}}\cdot\exp(-2\eta_{\mathrm{eff}}(H)x).
\end{align}
The factor $2$ results from the fact that the BLS intensity is proportional to the SAW intensity, which is again proportional to the squared SAW amplitude. We determine the effective damping by plotting the BLS-intensity as a function of the propagation length $x$ for each magnetic field in logarithmic representation as illustrated in Fig.\;\ref{Fig:Spacedep_PhononAbs}. The obtained effective damping rates $\eta_{\mathrm{eff}}$ are shown in Fig.\;\ref{Fig:AbsorptionRates}. The decay rate increases at the resonant coupling field with different magnitudes, indicating a non-reciprocal SAW-SW coupling~\cite{Kuss:Chiral:2022}, by 74\;\% at +11\;mT and 41\;\% at -9\;mT in comparison to off-resonant fields.
%The decay in SAW amplitude is increased up to 3.8\;µm$^{-1}$ at 11\;mT ( 2.1\;µm$^{-1}$ at -9\;mT). 
%
%

%%%%%%%%%%%%%%%%%%%%%%%%%%%%%%%%%%%%%%%%%%%%%%%%%%%%%%%%%%%%%%%%%%%%%%%%%%%%%%%%%%%%%%%%%%%%%%%%%%%%%%%%%%%%%%% Conclusion %%%%%%%%%%%%%%%%%%%%%%%%%%%%%%%%%%%%%
In summary, we demonstrated spatially resolved coherent interaction between phonons and magnons by micro-focused Brillouin light scattering experiments. By exploiting the shift in polarization of light scattered by magnons, we selectively detected the excitation of magnons and the absorption of phonons as a function of the applied magnetic field. We found that magnon and phonon signals interfere, demonstrating their coherence. By taking the coherent phase relation between SAW and SW into consideration, we formulated a phenomenological model for the expected BLS intensity, that we used to fit our data. %Finally, we characterized the nonreciprocal phonon magnon coupling efficiency by determining the magnetic field dependent phonon absorption. 
Our spatially resolved data shows that the SAW-SW interaction does not result in increased SW propagation length~\cite{Casals:Generation:2020}. This finding as well as the interference of phonons and magnons need to be considered for potential applications that rely on magnetoacoustically generated magnons or magnon-controlled phonon propagation. 

\begin{acknowledgments}
This work was funded by the Deutsche Forschungsgemeinschaft (DFG, German Research Foundation) – project number 492421737, the DFG  TRR 173 - 268565370 (project B01) and by the European Union within the HORIZON-CL4- 2021-DIGITAL-EMERGING-01 Grant No. 101070536 M\&MEMS.
\end{acknowledgments}

\clearpage

\bibliographystyle{apsrev4-2}
\bibliography{Bibliography}

\end{document}

% \maketitle
% \section{Supplemental Materials}

\title{\Large Supplementary Material \\ \large Coherent phonon-magnon interactions detected by micro-focused Brillouin light scattering spectroscopy}%

\author{Yannik Kunz} 
%\email{ykunz@rptu.de}
\affiliation{Fachbereich Physik and Landesforschungszentrum OPTIMAS, Rheinland-Pfälzische Technische Universität Kaiserslautern-Landau, 67663 Kaiserslautern, Germany}
\author{Matthias Küß}
\affiliation{Institute of Physics, University of Augsburg, 86135 Augsburg, Germany}
\author{Michael Schneider}
\author{Moritz Geilen}
\author{Philipp Pirro}
\affiliation{Fachbereich Physik and Landesforschungszentrum OPTIMAS, Rheinland-Pfälzische Technische Universität Kaiserslautern-Landau, 67663 Kaiserslautern, Germany}
\author{Manfred Albrecht}
\affiliation{Institute of Physics, University of Augsburg, 86135 Augsburg, Germany}
\author{Mathias Weiler}
\affiliation{Fachbereich Physik and Landesforschungszentrum OPTIMAS, Rheinland-Pfälzische Technische Universität Kaiserslautern-Landau, 67663 Kaiserslautern, Germany}

\maketitle

\subsection{Phenomenological description of Phonon-Magnon Coupling}
\label{AppendixPMCoupling}

The possibility of coupling ultrasonic waves with spin waves (SW) was proposed by Charles Kittel in 1958~\cite{Kittel1958}. The interaction mechanism invokes the strain dependence of the anisotropy field in the presence of spin-orbit coupling. Here, the surface acoustic wave (SAW) induces a time- and space-dependent driving field that interacts with the SW system. A semi-classical description was presented by Dreher \textit{et al.}~\cite{Dreher2012} by using an effective field approach, where the magnetic system is described by the Landau-Lifshitz-Gilbert equation
\begin{align}
\partial_t\mathbf{m}=-\gamma\mathbf{m}\times\mu_0\mathbf{H}_{\mathrm{eff}}+\alpha\mathbf{m}\times\partial_t\mathbf{m}.\label{Eq:LLG}
\end{align}
Here, $\gamma$ is the gyromagnetic ratio, $\alpha$ the Gilbert-damping parameter, $\mathbf{m}=\textbf{M}/M_\text{S}$ the unit vector of magnetization, and $\mu_0\mathbf{H}_{\mathrm{eff}}$ the effective magnetic field. The SAW enters Eq.\;\eqref{Eq:LLG} by inducing a periodic strain that contributes to the free enthalpy density $G^{\mathrm{tot}}$, which in turn governs the effective magnetic field $\mu_0\mathbf{H}_{\mathrm{eff}}$, which is determined by the vector differential of $G^{\mathrm{tot}}$ with respect to the components of $\mathbf{m}$
\begin{align}
\mu_0\mathbf{H}_{\mathrm{eff}}=-\nabla_{\mathbf{m}}G^{\mathrm{tot}}.
\end{align}
For the effective magnetic field we take into consideration the external magnetic field $\mu_0\textbf{H}_{\text{ext}}$, the anisotropy field $\mu_0\textbf{H}_{\text{ani}}$, the out-of-plane surface anisotropy field $\mu_0\textbf{H}_\text{k}$, the dipol- and the exchange fields. The solution of the Landau-Lifshitz-Gilbert equation is given in terms of the inverse magnetic susceptibility tensor $\bar{\chi}^{-1}$
\begin{align}
    \bar{\chi}^{-1}=&\frac{1}{M_\text{S}}\left(\begin{matrix}
\chi_{11}^I & \chi_{12}^I\\
\chi_{21}^I & \chi_{22}^I
\end{matrix}
    \right),\\
    \chi_{11}^I=&H\cos(\phi_0-\phi_H)+\frac{2A}{\mu_0M_\text{S}}k^2+M_\text{S}G_0-H_\text{k}+H_\text{ani}\cos^2(\phi_0-\phi_\text{ani})-i\frac{\alpha\omega}{\mu_0\gamma},\notag\\
    \chi_{12}^I=&-\chi_{21}^I=i\frac{\omega}{\mu_0\gamma},\notag\\
    \chi_{22}^I=&H\cos(\phi_0-\phi_H)+\frac{2A}{\mu_0M_\text{S}}k^2+M_\text{S}(1-G_0)\sin^2(\phi_0)-H_\text{k}+\notag\\
    &H_\text{ani}\cos(2(\phi_0-\phi_\text{ani}))-i\frac{\alpha\omega}{\mu_0\gamma},\notag\\
    \chi_{11}=&\frac{\chi^I_{22}}{M_\text{S}\cdot\text{det}(\bar{\chi}^{-1})}.
\end{align}
Here, $M_\text{S}$ is the saturation magnetization, $\phi_0, \phi_\text{H}$ and $\phi_\text{ani}$ are the angles between the die SAW wavevector $\mathbf{k}_\text{SAW}$ and the magnetization $\mathbf{m}$, external magnetic field $\mu_0H=\|\mu_0\mathbf{H}_\text{ext}\|$ and the in-plane anisotropy easy-axis, respectively. $A$ is the exchange constant, $H_\text{k}$ the out-of-plane surface anisotropy field, $G_0=(1-e^{-\|k\|d})/(\|k\|d)$ is a dipolar spin wave term~\cite{KalinikosSlavin1986} and $\omega$ the angular frequency.
The magnetoelastic contribution to the free enthalpy is expressed by
\begin{align}
G^{\mathrm{ela}}=&b_1[\varepsilon_{xx}m_x^2+\varepsilon_{yy}m_y^2+\varepsilon_{zz}m_z^2]\notag\\
&+2b_2[\varepsilon_{xy}m_xm_y+\varepsilon_{xz}m_xm_z+\varepsilon_{yz}m_ym_z],
\end{align} 
with $b_1$ and $b_2$ denote the magnetoelastic coupling coefficient. As the SAW causes a periodic deformation, the components $\varepsilon_{ij}$ become time- and space-dependent, thus inducing a small magnetic driving field $\mathbf{h}_{\mathrm{dr}}(\mathbf{r},t)$. Following the approach of Dreher \textit{et al.}~\cite{Dreher2012}, the driving field can be written as:
\begin{align}
\mathbf{h}_{\mathrm{dr}}(\mathbf{r},t)=\frac{2}{\mu_0M_{\mathrm{S}}}\begin{pmatrix}
b_2\varepsilon_{xz}\\
b_{1}\varepsilon_{xx}\sin(\phi_0)
\end{pmatrix}\cos(\phi_0).\label{Eq:hdr}
\end{align}
Via the driving magnetic field, the SAW can couple to the magnetic material and excite a SW in the latter. In the quasi-particle picture this process represents phonon-magnon coupling. A resonant interaction between the two systems occurs if energy and momentum are conserved in the scattering process:
\begin{align}
h f_{\mathrm{Magnon}} = h f_{\mathrm{Phonon}}\\
\hbar k_{\mathrm{Magnon}} = \hbar k_{\mathrm{Phonon}}
\end{align}
Thus, resonant coupling is found at the intersection of the dispersion relations of the SAW and the SW. As the dispersion relation of the latter depends on the magnitude and angle of the external magnetic field, the SAW transmission can be affected by choosing these correspondingly. This yields an angle- and magnetic field dependent phonon transmission. The nonreciprocal SAW-SW coupling is induced by the helicity mismatch effect~\cite{Xu2020, Kuess2020, Sasaki2017}, where the coupling efficiencies depend on the relative helicities of the SAW and the SW. The non-reciprocity can be further enhanced by using magnetic thin film systems with a nonreciprocal SW dispersion relation $f_\text{SW}(-k)\neq f_\text{SW}(+k)$. This can be achieved for instance by introducing Dzyaloshinskii–Moriya interaction (DMI) to the system and leading to an antisymmetric SW dispersion~\cite{Kuess2020}, the dipolar coupling of ferromagnetic bilayers~\cite{Kuess2021} or synthetic antiferromagnets~\cite{Kuess2022}.\\

\subsection{Experimental Methods}\label{ExpMeth}
\label{AppendixExpMeth}
To investigate the nonreciprocal phonon-magnon coupling we performed microfocused Brillouin light scattering experiments. In general, the coupling efficiency between SAWs and SWs is investigated using electrical methods, for example employing vector network analysis~\cite{Kuess2020, Xu2020, Weiler2011}. While this allows for direct characterization of the global SAW transmission, one loses the spatial resolution. Using optical investigation methods with a spatial resolution of around 300\;nm and being sensitive~\cite{Kargar2021} and selective~\cite{Geilen:Fully:2022} to both phonon and magnon signal, we are able to resolve the magnetoacoustic interaction and map the magnetic field and the space-dependent absorption of phonons and the excitation of magnons. We employ a microfocused BLS experiment with a 532\;nm CW single mode laser, which is focused using a microscope objective with a numerical aperture of NA$\;=\;$0.75, limiting the maximum resolvable wave vector to $k_\text{max}\approx16$\;rad/µm. A schematic depiction of the measurement technique and the sample is given in Fig.\;1 a) of the main text. The frequency shift of the inelastic scattered light is analyzed by a Tandem-Fabry-Pérot Interferometer (TFPI). When a rotatable plate $\lambda$/2 is placed in front of the polarization-sensitive TFPI, selective detection of phonons and magnons is enabled as the light scattered by magnons is shifted in polarization by 90° with respect to the phonon signal, similar to the magneto-optical Kerr effect (MOKE).\\
We investigated the phonon-magnon interaction on a 400\;µm long and 200\;µm wide ferromagnetic Co\lc{40}Fe\lc{40}B\lc{20}(10\;nm)/Si\lc{3}N\lc{4}(3\;nm) layer, which was deposited by magnetron sputtering deposition at room temperature on a piezoelectric Y-cut Z-propagation LiNbO\lc{3} substrate, that supports the Rayleigh-type SAW~\cite{Morgan2007}. To excite the acoustic wave, we use sets of IDTs made of Ti(5\;nm)/Al(70\;nm) with a periodicity of 6.8\;µm and 30 finger pairs, to which a microwave voltage is applied. The oscillating electric field between the fingers of the IDT induces a periodic strain in the LiNbO\lc{3}. The Co\lc{40}Fe\lc{40}B\lc{20} layer and the Ti/Al-IDTs were deposited using magnetron sputtering deposition and electron beam evaporation, respectively. For more details on the sample fabrication, please see Ref.\;~\cite{Kuess2020}.

\subsection{Magnetic properties of Co\lc{40}Fe\lc{40}B\lc{20}}
\label{AppendixFMR}
\begin{figure}[h]
\centering
%\centerline{\include{Graphs/FMR.eps}}
\includegraphics[scale = 1]{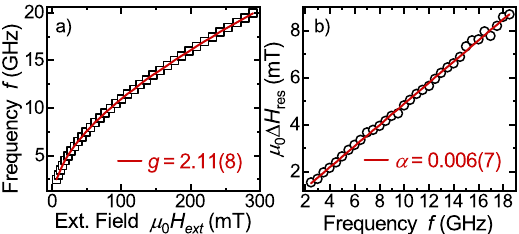}
\caption{The ferromagnetic resonance (FMR) of Co\lc{40}Fe\lc{40}B\lc{20}. Panel a) shows the dependence of the FMR frequency $f$ of the external magnetic field $\mu_0H$, fitted with Kittels' equation. Panel b) shows the resulting inhomogeneous Linewidth $\mu_0\Delta H_{\mathrm{res}}$, the slope of the fitted linear is proportional to the Gilbert-damping parameter $\alpha$.}
\label{Fig7}
\end{figure}
In the first step of our investigation we determined the magnetic properties of the Co\lc{40}Fe\lc{40}B\lc{20}-layer by using a reference sample, that was fabricated under the same conditions as the sample investigated by µBLS. For this, well established broadband ferromagnetic resonance spectroscopy~\cite{MaierFlaig2018} was employed, providing the g-factor $g=2.11(8)$, Gilbert damping parameter $\alpha=0.006(7)$, saturation magnetization $\mu_0M_{\mathrm{S}}=1.287$\;T and the anisotropy field $\mu_0H_{\mathrm{ani}}=-1.46$\;mT (see Fig.\;\ref{Fig7}). Our experimental finding on the magnetic parameters of Co\lc{40}Fe\lc{40}B\lc{20} is in agreement with the values of the literature and previous measurements in other publications~\cite{Kuess2020, Cho2013}. We note, that Co\lc{40}Fe\lc{40}B\lc{20} possesses a comparably high Gilbert damping parameter, thus giving rise to a low magnon lifetime and decay length $\xi_\text{SW}$ by~\cite{Stancil2009}
\begin{align}
    %\xi_\text{SW}=\frac{1}{\alpha2\pi f_\text{SW}}\cdot\frac{\partial f_\text{SW}}{\partial k},
    \xi_\text{SW}=&\frac{2\pi}{\alpha\gamma\mu_0(H_0+M_\text{S}(1-g(k)\cos^2(\varphi)/2))}\cdot\frac{\partial f_\text{SW}}{\partial k},\label{Eq:SW_DecayLength}\\
    &g(k)=1-\frac{1-\exp(-\|k\|d)}{\|k\|d},    
\end{align}
of $\xi_{\mathrm{SW}}\approx1.81$\;µm, at $f=5$\;GHz, $\mu_0H=10$\;mT in Damon-Eshbach geometry. In Eq.\;\eqref{Eq:SW_DecayLength}, $f_\text{SW}$ denotes the SW frequency given by the dispersion relation Eq.\;(2) of the main text.

\bibliographystyle{apsrev4-2}
\bibliography{BibliographySM}